\documentclass[nobibnotes,nofootinbib,aps,pra,reprint, showpacs, superscriptaddress]{revtex4-2}

\usepackage[bottom]{footmisc}
\usepackage{epsfig,graphics,graphicx}% Include figure files
\usepackage{dcolumn}% Align table columns on decimal point
\usepackage{bm}% bold math
\usepackage{color}
\usepackage{amsmath,amssymb,amsthm}
\usepackage{centernot}
\usepackage{bbold}
\usepackage{color,soul}
\usepackage{mathtools}
\usepackage{fullpage}
\usepackage{units}
\usepackage[usenames,dvipsnames]{xcolor}
\usepackage{braket}
\usepackage{indentfirst}
\usepackage{lipsum}
\usepackage[export]{adjustbox}
\usepackage{enumerate}
\usepackage{appendix}
\usepackage{gensymb}

\usepackage[utf8]{inputenc}
\usepackage[T1]{fontenc}
\usepackage{dsfont}

\usepackage{mathptmx}
\usepackage{amssymb}
\usepackage{amsmath}
\usepackage{latexsym}
\usepackage{amsfonts}
\usepackage{float}
\usepackage{hyperref}
\hypersetup{pdfpagemode=UseNone}
\newtheorem{theorem}{Theorem}
\newtheorem{definition}[theorem]{Definition}

\newtheorem{lemma}[theorem]{Lemma}
\newtheorem{proposition}[theorem]{Proposition}

\newcounter{rem}
\def\>{\rangle}
\def\<{\langle}

\renewcommand{\rho}{\varrho}

\def\textbf#1{{\bf #1}}

\def\beq{\begin{equation}}
\def\eeq{\end{equation}}
\def\beqa{\begin{eqnarray}}
\def\eeqa{\end{eqnarray}}
\def\eea{\end{array}}
\def\bea{\begin{array}}
\newcommand{\bei}{\begin{itemize}}
\newcommand{\eei}{\end{itemize}}
\newcommand{\bee}{\begin{enumerate}}
\newcommand{\eee}{\end{enumerate}}
\def\bep{\begin{proposition}}
\def\eep{\end{proposition}}
\def\bel{\begin{lemma}}
\def\eel{\end{lemma}}
\def\bet{\begin{theorem}}
\def\eet{\end{theorem}}
\def\bed{\begin{definition}}
\def\eed{\end{definition}}

\begin{document}

\title{Equilibration of Isolated Systems: investigating the role of coarse-graining on the initial state magnetization}

\author{Gabriel Dias Carvalho}
\email{gdc@poli.br}
\affiliation{Física de Materiais, Universidade de Pernambuco, 50720-001, Recife, Pernambuco, Brazil}
\affiliation{Instituto de Física, Universidade Federal Fluminense, Niterói, Rio de Janeiro 24210-346, Brazil}

\author{Luis Fernando dos Prazeres}
%\email{}
\affiliation{Instituto de Física, Universidade Federal Fluminense, Niterói, Rio de Janeiro 24210-346, Brazil}

\author{Pedro Silva Correia}
\email{pscorreia@uesc.br}
\affiliation{Departamento de Ciências Exatas, Universidade Estadual de Santa Cruz, Ilhéus, Bahia 45662-900, Brazil}

\author{Thiago R de Oliveira}
%\email{}
\affiliation{Instituto de Física, Universidade Federal Fluminense, Niterói, Rio de Janeiro 24210-346, Brazil}

\date{\today}

%%%%%%%%%%%%%%%%%%%%%%%%%%%%%%%%%%%%%%%%%%%%%
%%%%%%%%%%%%%Abstract%%%%%%%%%%%%%%%%%%%%%%%%
%%%%%%%%%%%%%%%%%%%%%%%%%%%%%%%%%%%%%%%%%%%%%

\begin{abstract}

Many theoretical and experimental results show that even isolated quantum systems evolving unitarily may equilibrate, since the evolution of some observables may be around an equilibrium value with negligible fluctuations most of the time. There are rigorous theorems giving the conditions for such equilibration to happen. In particular, initial states prepared with a lack of resolution in the energy will equilibrate. We investigate how equilibration may be affected by a lack of resolution, or coarse-graining, in the magnetization of the initial state. In particular, for a chaotic spin chain and using exact diagonalization, we show that the level of equilibration of an initial state with a coarse, not well-defined magnetization is different from the level of an initial state with well-defined magnetization. This difference will depend on the degree of coarse-graining and the direction of magnetization. We also analyze the time for the system to reach equilibrium, showing good agreement with theoretical estimates and with some evidence that less resolution leads to faster equilibration. Our study highlights the crucial role of initial state preparation in the equilibration dynamics of quantum systems and provides new insights into the fundamental nature of equilibration in closed systems. 
\end{abstract}


\maketitle

%%%%%%%%%%%%%%%%%%%%%%%%%%%%%%%%%%%%%%%%%%%%%%%%%%%
%%%%%%%%%%%%%%%%%%%%%%%%%%%%%%%%%%%%%%%%%%%%%%%%%%%
%%%%%%%%%%%%%%%%%%%%%%%%%%%%%%%%%%%%%%%%%%%%%%%%%%

%%%%%%%%%%%%%%%%%%%%%%%%%%%%%%%%%%%%%%%%%%%%%%%%%%
%%%%%%%%%%%%%%Introduction%%%%%%%%%%%%%%%%%%%%%%%
%%%%%%%%%%%%%%%%%%%%%%%%%%%%%%%%%%%%%%%%%%%%%%%%5

\section{Introduction}

One of the basic assumptions of thermodynamics is that systems will reach thermal equilibrium if left alone after some perturbation. It is also one of the main foundational problems of statistical physics; to show that equilibration occurs from microscopic principles. One main approach is to consider that the system is not isolated and that interaction with the external environment will lead to thermal equilibrium.

A different approach, that has gained a lot of attention, is to show that even isolated systems described by pure states may reach thermal equilibrium when one is monitoring only a few observables \cite{ReviewNatureEisert} \footnote{Actually, this approach was proposed
already by Boltzmann \cite{Lebowitz93, Lebowitz07} and extended to the quantum scenario by von Neumann \cite{Goldstein10}. There is also some discussion if this equilibration process is, in fact, equivalent to the decoherence of open systems; see \cite{Argentinos} for
a positive view and \cite{Schlosshauer05} for a negative one.}. Since the dynamic is unitary, the expectation value of an observable should never stop oscillating, but such oscillations may be very small most of the time and the system appears as in equilibrium. Rigorous theorems show that to equilibrate, initial states must be in a superposition containing a large number of energy eigenvectors, which is equivalent to states prepared with an apparatus that lack resolution in energy \cite{Reimann2008, Popescu2009}.
It can be argued that since for local Hamiltonians, the distance between energy levels decays exponentially with the number of particles, it is almost impossible to prepare an initial state with just a few eigenstates. Thus, we should expect that most states prepared in the lab, or nature, will equilibrate if left isolated. From a more mathematical point of view, it is possible  
to show that when randomly chosen using a Haar measure, most of the states will equilibrate 
(see theorem 2 in \cite{Popescu2009}).

While in terms of energy it is clear when an initial state will equilibrate, there are no general or rigorous results about other physical properties, such as magnetization,
correlations, or entropy. The few exceptions are \cite{Brandao}, where the authors show that states with exponentially decaying correlations equilibrate, and \cite{WilmingEisert}, where it is shown that product states equilibrate when the system's energy eigenstates at finite energy density have a sufficient amount of entanglement. There are many numerical works, for particular spins chains with an initial product state and well-defined local magnetization, that explore how the changes in the parameters of $H$ (a quench) affect equilibration and thermalization \cite{Polkovniov11, Gogolin16}, and a few works looking explicitly at the role of the properties of the initial states \cite{Banuls11, Lea2013, Rigol2012, Fagotti2016}.

In this work, we analyze how the lack of resolution in the magnetization of the initial state affects equilibration.
In particular, we study two different situations: (a) we do not have the complete spatial resolution to prepare the initial state, namely, we know the number of spins pointing up but do not know their positions on the lattice; (b) we do not know how many spins are pointing up and also where they are. In case (a) we have a well-defined global but not local magnetization. In case (b) neither a global nor a local well-defined magnetization. This context of lack of resolution is what justifies our use of the term coarse-graining \cite{CrisCG}. Thus, we will consider as our initial state a coherent superposition in the local basis of magnetization, which is not the same as the energy basis. We show that this addition of coherences will be relevant in the process of equilibration.  As expected, this is because coarse-grained states are not a superposition of a few energy eigenstates. 

Note that the effects of lack of resolution are already present in the experimental literature. In \cite{PinheiroExp} the authors analyze an experiment with p-bosons in an XYZ quantum Heisenberg model and discuss experimental imperfections in the system's preparation, detection, and manipulation, emphasizing the rich physics involved. In \cite{atomlattice} a spin-lattice is emulated in a lattice where individual atoms are trapped in each potential minimum. The measuring detector is such that it cannot resolve light coming from each atom individually.

In the present article we demonstrate a relationship between the degree of equilibration and the degree of coarse-graining on the initial state. The results, however, do not indicate that the degree of equilibration can be completely suppressed, in the thermodynamic limit, by the level of accuracy, they only indicate that the propensity for equilibration is related to this accuracy for finite systems. We also briefly study the time scale for the equilibration to happen and compare with results in \cite{ThiagoNJ}. There, the authors describe the equilibration of an observable as the dephasing of the complex amplitudes of the energy gaps, estimating theoretically from heuristic arguments the equilibration time in terms of the dispersion of the distribution of energy gaps. 

Our article is organized as follows: In section \ref{sec:isosyseq} we discuss some general features of equilibration of isolated systems. In section \ref{sec:isi} we discuss our system of interest, including the Hamiltonian, the measured observable, and the initial state. Our initial states were explained and the extent of our ignorance was clarified. In section \ref{sec:decg} we present our main result, regarding the effective dimension of the system and the coarse-graining. In section \ref{sec:eqcg} we compare our equilibration times with values obtained via the procedure in \cite{ThiagoNJ}. 

\section{Equilibration of isolated systems}
\label{sec:isosyseq}

Consider an initial state $|\psi_0\> = \sum_{k} c_k |E_k \> $ evolving under $H= \sum_{k} E_k |E_k\>\<E_k|$ and lets monitor an observable $\hat{A}$ by measuring its expectation value $A(t)=\< \hat{A}(t)\>$. As the system is isolated and evolving unitarily, we expect $A(t)$ to oscillate and never reach a stationary state. However, it may oscillate very close to an equilibrium value most of the time, such that one is not able to distinguish it from an equilibrium; we have probabilistic equilibration. One way to quantify this is to look at the time average of the fluctuation; if it is small, the system is very close to equilibrium most of the time. It is possible to show that \footnote{There are many variations of this result. The original one is \cite{Reimann2008}, and the one we use with the operator norm was given in \cite{Short11}. It is also possible to consider reduced density operators instead of observables as in \cite{Popescu2009}. For a review of these results see \cite{Gogolin16}.}
\begin{equation}
\overline{(A(t)-\overline{A})^2} \leq ||A|| / d_{eff},
\label{equilibr}
\end{equation}
with $\overline{A}$ the infinite-time average of $A(t)$, $||A||$ the operator norm of $\hat{A}$ and
\begin{equation}
d_{eff}=\frac{1}{\sum |c_k|^4},
\label{deff}
\end{equation}
known as the effective dimension. $d_{eff}$ is a measure of how many energy eigenstates participate in the initial state $\ket{\psi_0}$ or how much of the Hilbert space is explored during the time evolution; hence the name effective dimension ($d_{eff}$ is also known as the participation ratio, a measure of the localization of a state in the energy basis). The result above is valid for Hamiltonians without degeneracies and "degenerate gaps": $E_k-E_l=E_m-E_n$ only if $E_k=E_m; E_l=E_n$ or $E_k=E_l; E_m=E_n$. While it is generally expected and numerically found that such conditions are true for systems of interacting particles, the results may be generalized for systems with some, not exponentially large, degeneracies and some "degenerated gaps" \cite{Short12} \cite{ReimannNJ2012}.

These analytical and general results give the mathematical ingredients needed for a system to equilibrate: a $H$ without many degenerated gaps and an initial state prepared with a lack of resolution in the energy. It is expected that, for systems with local interactions, it is almost impossible to prepare in the lab a $|\psi_0\>$ with low $d_{eff}$, since the distance between the energy levels becomes much smaller than any actual experimental resolution (for more details see sec. 2.1 of \cite{ReimannNJ2012} and \cite{Gallego} for quantification of the resources need to prepare an initial state that does not equilibrate). However, it is still unclear which other physical properties of a system and state are important for equilibration. There are a few analytical results for particular cases that equilibrate: (1) states with exponentially decaying correlations \cite{Brandao}, and (2) product states when the system's energy eigenstates at finite energy density have a sufficient amount of entanglement \cite{WilmingEisert}. A few negative results also show that some simple initial product states never equilibrate \cite{Wilming22}. On the other hand, many numerical results study how the Hamiltonian parameters affect equilibration and thermalization,
mostly considering initial product states \cite{Polkovniov11, Gogolin16}, and some analyzing the effects of different initial states
\cite{Banuls11, Lea2013, Rigol2012, Fagotti2016}.

Instead of considering the resolution in the preparation of the system in terms of its energy, we will consider in terms of the system's magnetization. 

\section{Initial States and Coarse-Graining}
\label{sec:isi}

We will consider a one-dimensional spin lattice with $n$ from $2$ to $16$ spins that evolve with a Hamiltonian based on the usual quantum Heisenberg XYZ model with an external magnetic field in $z$ and open boundary conditions \cite{Naoto, Nozawa}. We also add a next-nearest-neighbor interaction, normally used to break integrability and avoid a large number of degenerate gaps \cite{Lea2012, Lea1}. The Hamiltonian is thus given by: 
\begin{equation}
\label{Hamiltonian}
H =  \sum_{i=1}^{n} J \, ( S_{i}^{x} S_{i + 1}^{x} + J_{y} \, S_{i}^{y} S_{i + 1}^{y} + J_{z} \, S_{i}^{z} S_{i + 1}^{z} +   S_{i}^{z} S_{i + 2}^{z} + h_{z} \,  S_{i}^{z}) .
\end{equation}
We set $J = 1$ and $\hbar = 1$ and $S_{i}^{x,y,z}$ are spins $1/2$ operators acting at site $i$. All the coupling parameters $J_{y}, J_{z}$ and $h_{z}$ are assumed to be positive. The measured observable is the magnetization in $z$ direction, given by
\begin{equation}
M_{z}(n) = \sum_{i=1}^{n} \mathbb{1}^{i-1} \otimes \sigma_{z} \otimes \mathbb{1}^{n-i},
\label{observable}
\end{equation} 
with $\mathbb{1}^{i-1}$ the identity operator acting on the first $i-1$ spins and $\sigma_{z}$ a Pauli matrix. 

We want to study how the resolution in the preparation of a magnetic state affects the equilibration. As mentioned, most studies consider a full magnetic product state (or anti-ferro):
\begin{equation}
	\ket{\psi_{0}} = \ket{ \uparrow . . . \uparrow \uparrow \uparrow }.
	\label{psi0nocg}
\end{equation}
To prepare such a state, for large systems where equilibration may occur, we need a very high resolution in the measure of the total spin, or be able to measure each spin locally.

The concept of coarse-graining is incorporated into our framework through the composition of our initial state. We consider a superposition where the number of states involved depends on our ability to resolve such a state. Due to our limited ability to resolve the initial state, we make the following assumptions: (a) we can resolve the total magnetization of the initial state, but not the direction of individual spins; and (b) we can only resolve that the total magnetization is larger than a given value. Therefore, we consider our initial state as a superposition of spin configurations that can not be distinguished within the apparatus resolution.

To illustrate case (a), consider a system of 3 spins with total magnetization corresponding to two spins pointing up. We can not distinguish between the states $|\downarrow  \uparrow  \uparrow\> $, $|\uparrow  \downarrow  \uparrow\>$ and $|\uparrow  \uparrow  \downarrow\>$. Then, our initial state is $|\downarrow  \uparrow  \uparrow\> + |\uparrow  \downarrow  \uparrow\> + |\uparrow  \uparrow  \downarrow\>$. Note that in case (a) we have a well-defined total magnetization. The states that fall into this category are analogous to Dicke states \cite{Dicke}. To illustrate case (b), consider a system of 3 spins where we can not distinguish between state $|\uparrow  \uparrow  \uparrow\> $ and the states with one spin flipped, as $|\downarrow  \uparrow  \uparrow\> $. Then, our initial state is $|\uparrow  \uparrow  \uparrow\> + |\downarrow  \uparrow  \uparrow\> + |\uparrow  \downarrow  \uparrow\> + |\uparrow  \uparrow  \downarrow\>$. This is equivalent to a coarse-graining measure of the total magnetization in the $\hat{z}$ direction.

In both cases, we utilize coherent sums in the form of pure states to describe our initial states. This choice is driven by two main considerations: the first is our assumption that the system is completely isolated, which means there is no leak of information for the environment or the system isn't part of a larger entangled system. The second is our desire to avoid introducing probabilities by hand. Had we chosen statistical mixtures as initial states, probabilities would have been present from the start.

To quantify the coarse-graining, we can define a parameter $k$, which is related to the number of flipped-down spins. In the $k = 0$ situation we have, in both cases, full resolution in the measure of the magnetization in the $\hat{z}$ direction (no coarse-graining at all) and can ensure that all spins are pointing in the up-direction (Eq. \ref{psi0nocg}). In the situation in which $k \ne 0$,  all the ket-states with the number of pointing down spins equal (case (a)) and equal or less (case (b)) than $k$  participate in the sum. For example, let us consider the case where $n = 6$ and $k = 2$. Our initial states would be given by
\begin{equation}
\ket{\psi_{0}} = \ket{\uparrow \uparrow \uparrow \uparrow \downarrow \downarrow} + p.p.,
\label{psi0cg}
\end{equation}
for case (a), and
\begin{equation}
\ket{\psi_{0}} = \ket{\uparrow \uparrow \uparrow \uparrow \uparrow \uparrow } + \ket{ \uparrow \uparrow \uparrow \uparrow \uparrow \downarrow } + \ket{ \uparrow \uparrow \uparrow \uparrow \downarrow \downarrow} + p.p.,
\label{psi0cgc}
\end{equation}
for case (b), with "p.p." meaning "possible permutations". For simplicity, we choose to omit the normalization factors in the discussion above. The parameter $k$ controls the resolution, or the amount of coarse-graining, with greater $k$ meaning less resolution and thus more states participating in the superposition of the initial state. From now on, in the context of case (a), the initial state with $k$-flipped-down spins and their permutations will be called the $k$-flipped initial state. In the context of case (b), the superposition will also include states with until $k$-flipped-down spins and their permutations.

The main message of this section: due to our lack of resolution, we have to add terms in $\ket{\psi_0}$, transforming it into a sum of possible states. The states that will compose the sum will be determined by the two types of coarse-graining: (a) the spatial one, where we know the global magnetization but do not know where the flipped-down spins are located. In this case, all possible permutations with $k$-flipped-down spins must compose the superposition; (b) the magnetization spectrum coarse-graining together with the spatial coarse-graining. In this case, we only know a magnetization lower bound, and all possible permutations with a $k$-flipped-down spins, or less, must compose the superposition.

\section{Effective Dimension and Coarse-graining}
\label{sec:decg}

One should not expect to have equilibration for small systems, since thermodynamics and statistical
physics only predict it for macroscopic systems. This is consistent with Eq. \ref{equilibr} since the upper bound can only be small for large systems. One should study the behavior of the average fluctuations with the number $n$ of particles; as in textbook statistical mechanics, where one shows that the relative fluctuation decays with $\sqrt{n}$.
Here we will analyze the behavior of $d_{eff}$ with $n$, using exact diagonalization, as it is
an upper bound on the average fluctuation: a large increase of effective dimension guarantees equilibration of any observable whose operator norm does not increase faster than $d_{eff}$ with $n$. Thus, we will look at how $d_{eff}$ scales with the system size $n$. For non-integrable systems, it is expected that $d_{eff}$ scales exponentially with $n$ \footnote{One may argue that a polynomial scaling may also be enough for equilibration of local observables and this is expected for some models that can be mapped in free fermions \cite{Venuti}.}.

\begin{figure}[!th]
\includegraphics[scale=0.8]{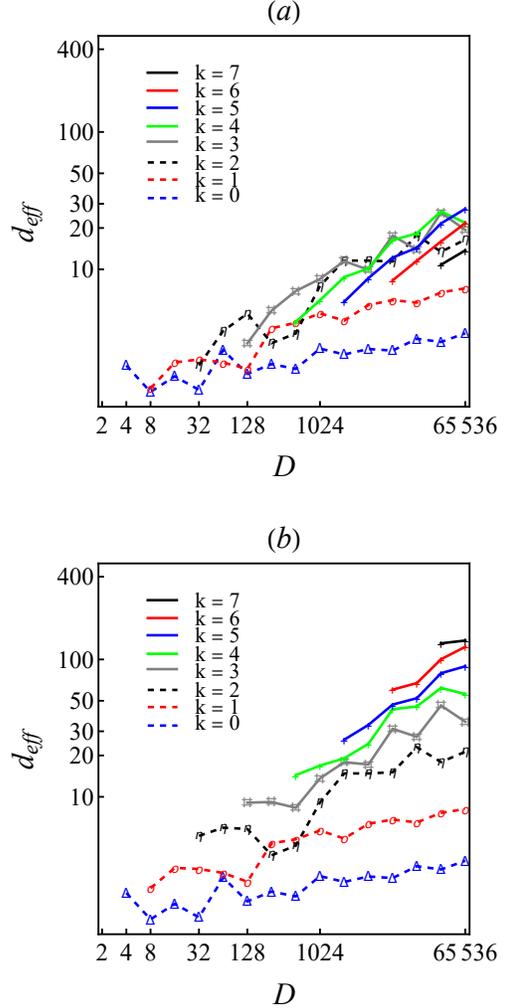} 
\caption{\small Effective dimension for case (a) and (b) as a function of the dimension $D = 2^n$ of the Hilbert space for different $k$-flipped initial states (intensity of the coarse-graining description in the $z$ direction magnetization) for Hamiltonian \ref{Hamiltonian}, with $J_{y} = 1.4 \, s^{-1}$, $J_{z} = 0.5 \, s^{-1}$ and $h_{z} = 0.01 \, s^{-1}$.}
\label{EdxD}
\end{figure}

To analyze the scaling of $d_{eff}$ with $n$, in Fig. \ref{EdxD} we show, in a log-log scale, the effective dimension of the coarse-grained initial state as a function of total Hilbert space dimension $D=2^{n}$ for spins chains with $n$ from 2 to 16 spins. Fig. \ref{EdxD}(a) is for case (a) and Fig. \ref{EdxD}(b) is for case (b). In both figures, the different curves are for initial states with different levels of coarse-graining, given by the $k$. For non-integrable systems, if equilibration is present, we should have $d_{eff} \sim \exp(n)$ and therefore $d_{eff} \sim D$. Thus to have equilibration we expect a linear scaling of $d_{eff}$ with $D$ in a log-log scale. As the dimension of the system increases, a large slope of the line means a stronger exponential decay; the fluctuation decreases faster with $n$. Note however that $d_{eff}$ gives only an upper bound on the size of the fluctuations of observables (Eq. \ref{equilibr}); it may happen that for some observables the equilibration occurs even when $d_{eff}$ is small.

Notice that for the $0$-flipped initial state, the effective dimension is almost constant in both (a) and (b) graphs, which means a small predilection for equilibration when compared to other $k$ values. The behavior of the graph then suggests that the coarse-graining induces equilibration. The greater the lack of resolution, the greater the trend of equilibration. As we will show in a more quantitative analysis, the results in Fig. \ref{EdxD} do not indicate that initial states with high accuracy do not exhibit equilibration in the thermodynamic limit, they only indicate that the propensity for equilibration depends on such accuracy for finite systems. Note that case (b) has more coarse-graining since it also includes a lack of knowledge in total magnetization. Thus, its initial states contain more states in the superposition and it is expected to have a larger $d_{eff}$; this can be seen in the larger values for $d_{eff}$ in Fig. \ref{EdxD}(b) when compared to Fig \ref{EdxD}(a). The curves in Fig. \ref{EdxD} have their beginnings for values of $D$ in which $n > 2k$. The reason is the fact that the set of states is symmetric concerning $k = \frac{n}{2}$. For example, a state of three spins with two spins up and one down is equivalent (same magnetization modulus) to a state with one spin down and two spins up. 

\begin{figure*}[!th]
\includegraphics[scale=1.0]{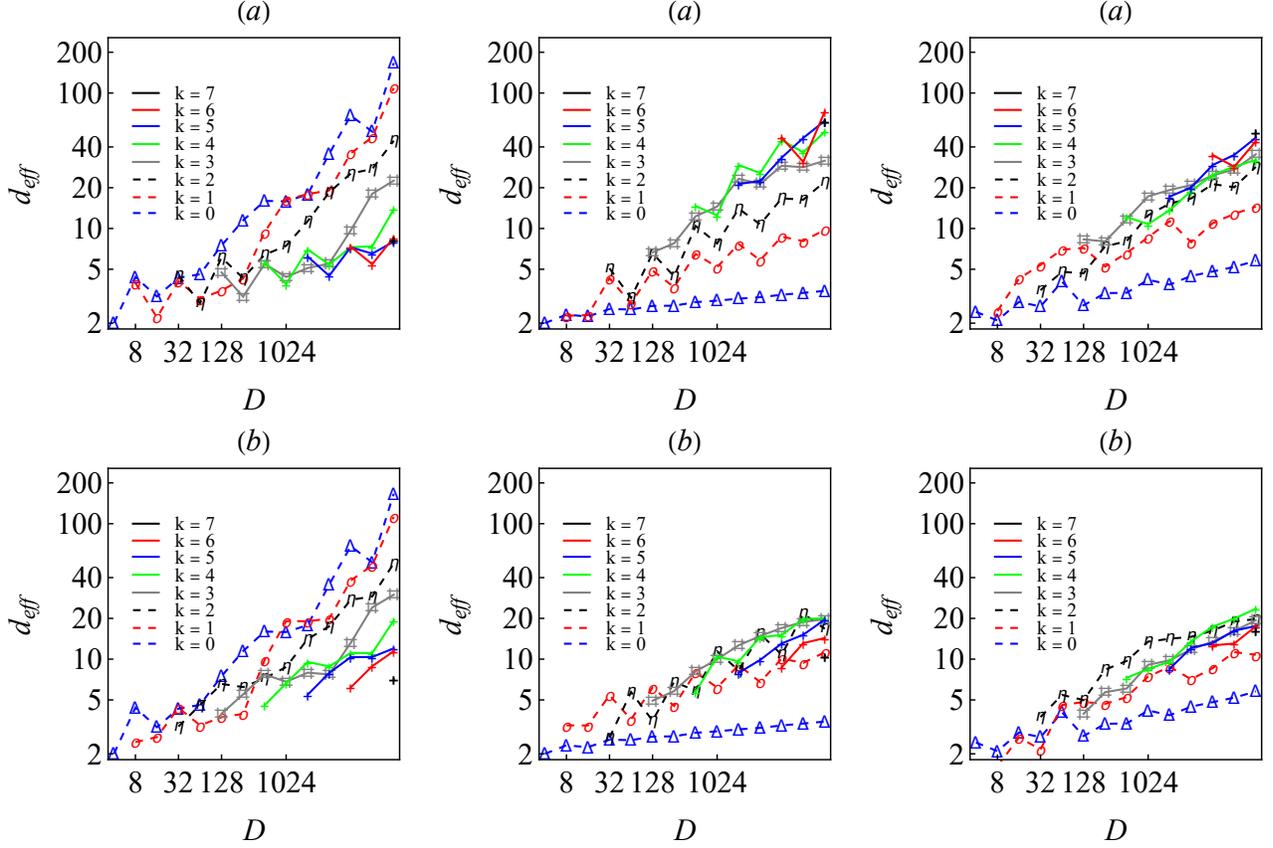} 
\caption{\small  Effective dimension for case (a) and (b) as a function of the dimension $D = 2^n$ ($n$ from $2$ to $15$) of the Hilbert space for different $k$-flipped initial states with lack of resolution in magnetization in the $x$ (first column), the $y$ (second column) and $(\theta,\phi)=(30\degree,90\degree)$ (third column) directions for Hamiltonian \ref{Hamiltonian}, with $J_{y} = 1.4 \, s^{-1}$, $J_{z} = 0.5 \, s^{-1}$ and $h_{z} = 0.01 \, s^{-1}$.}
\label{FigAppendix}
\end{figure*}

To check the behavior for other directions than $z$, in Fig. \ref{FigAppendix} we show similar plots of Fig. \ref{EdxD} for a coarse-graining of the magnetization in the $x$ (first column), $y$ directions (second column) and the direction given by the polar and azimuthal angles $\theta=30\degree$ and $\phi=90\degree$ (third column). In the first row are shown the effective dimensions for case (a) and in the second row are the effective dimensions for case (b). Inspecting the graphs, it is clear that the link between equilibration and resolution is present in directions other than the $z$ direction: coarse-graining affects the propensity for equilibration. For directions $z$, $y$ and $(\theta,\phi)=(30\degree,90\degree)$ coarse-graining increases equilibration, while in the $x$ direction it decreases. We suspect that the anisotropy and the Hamiltonian parameters are related to the directions that present a greater tendency toward equilibration.

\begin{figure*}[!th]
\includegraphics[scale=1.02]{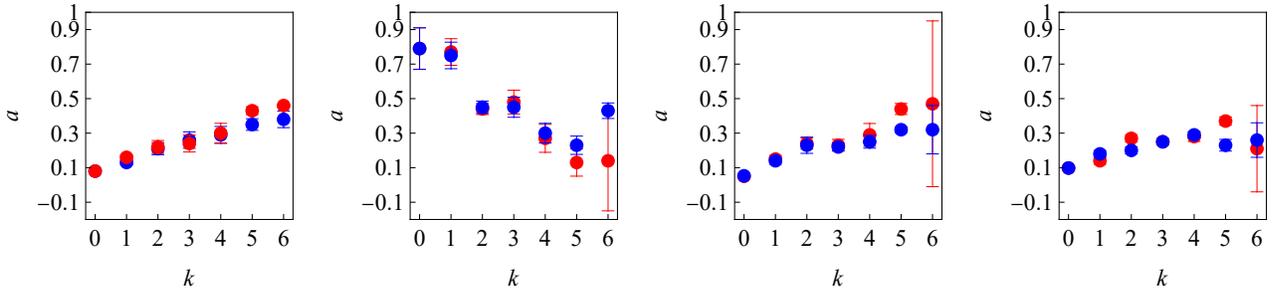} 
\caption{\small Slope of linear fits of the points in the graph of effective dimension versus $D = 2^n$ as a function of $k$ for case (a) (red points) and case (b) (blue points) for directions, from left to right, $z$, $x$, $y$ and given by the angles $(\theta = 30\degree, \phi = 90\degree)$. Except for direction $x$, the greater the lack of resolution of our initial state description, the greater the slope and the tendency to equilibrate. Note that, the number of points used in the fitting decreases with k, with $k=6$ having 4 points for direction $z$ and 3 points for the others, and thus the statistical significance of these values also decreases.}
\label{kxa}
\end{figure*}

For a more quantitative analysis, we fit the curves in Figs. \ref{EdxD} and \ref{FigAppendix} with the function $b \, D^a$, except the $7$-flipped curve, since it only has two points. In the log-log scale, the corresponding function represents linear fits with $a$, the angular coefficient, or the slope, of the lines. Fig. \ref{kxa} shows the slope coefficient $a$ versus $k$. The important point here is that the greater the number of flipped-down spins (in other words, the coarse-graining), the greater the slope of the linear fit in Figs. \ref{EdxD} and \ref{FigAppendix} and the tendency towards equilibration, except for the $x$ direction, case in which the lack of resolution seems to disfavor the tendency towards equilibration. This confirms the qualitative behavior of Figs. \ref{EdxD} and \ref{FigAppendix}. In the $z$, $y$ and $(\theta = 30\degree, \phi = 90\degree)$  directions, note that $a$ gives how fast the fluctuations decay with system size, and therefore for finite $a$ any system will equilibrate if large enough. In those directions, the coarse-graining seems to favor equilibration already for smaller systems. Including the $x$ direction, we can say that the coarse-graining already affects equilibration for smaller systems.

%Notice that $1 > a > 0$ for all $k$, which brings the discussion of the multifractality of eigenstates \cite{fractal}: an eigenstate is multifractal when it is extended but covers only a finite fraction of the available physical space. 

As mentioned before, $d_{eff}$ directly measures how many energy eigenstates contribute to the initial pure state. To see how our coarse-grained states behave in terms of the energy eigenstates, we look at the weighted energy distribution of the states \cite{Lea1}. This distribution is often called the Local Density of States (LDOS) or strength function, given by
\begin{equation}
P(E)= \sum_k |c_k|^2 \delta(E-E_k).
\end{equation}
Note that $d_{eff}$ is related to $P(E)$; in fact, it is the "purity" of the distribution $P(E)$. As $P(E)$ is more distributed in the spectrum, $d_{eff}$ is larger. The LDOS has been studied in many models and realistic systems with two-body interactions. Usually, the density of states is Gaussian and the initial state can only reach a high level of delocalization if its average energy is in the middle of the spectrum (see \cite{Lea1} for a review of some models and random Hamiltonians). We will now study the LDOS for the initial state with coarse-graining in the $z$ direction to illustrate the relation between the LDOS and $d_{eff}$.

\begin{figure*}[!th]
\includegraphics[scale=1.02]{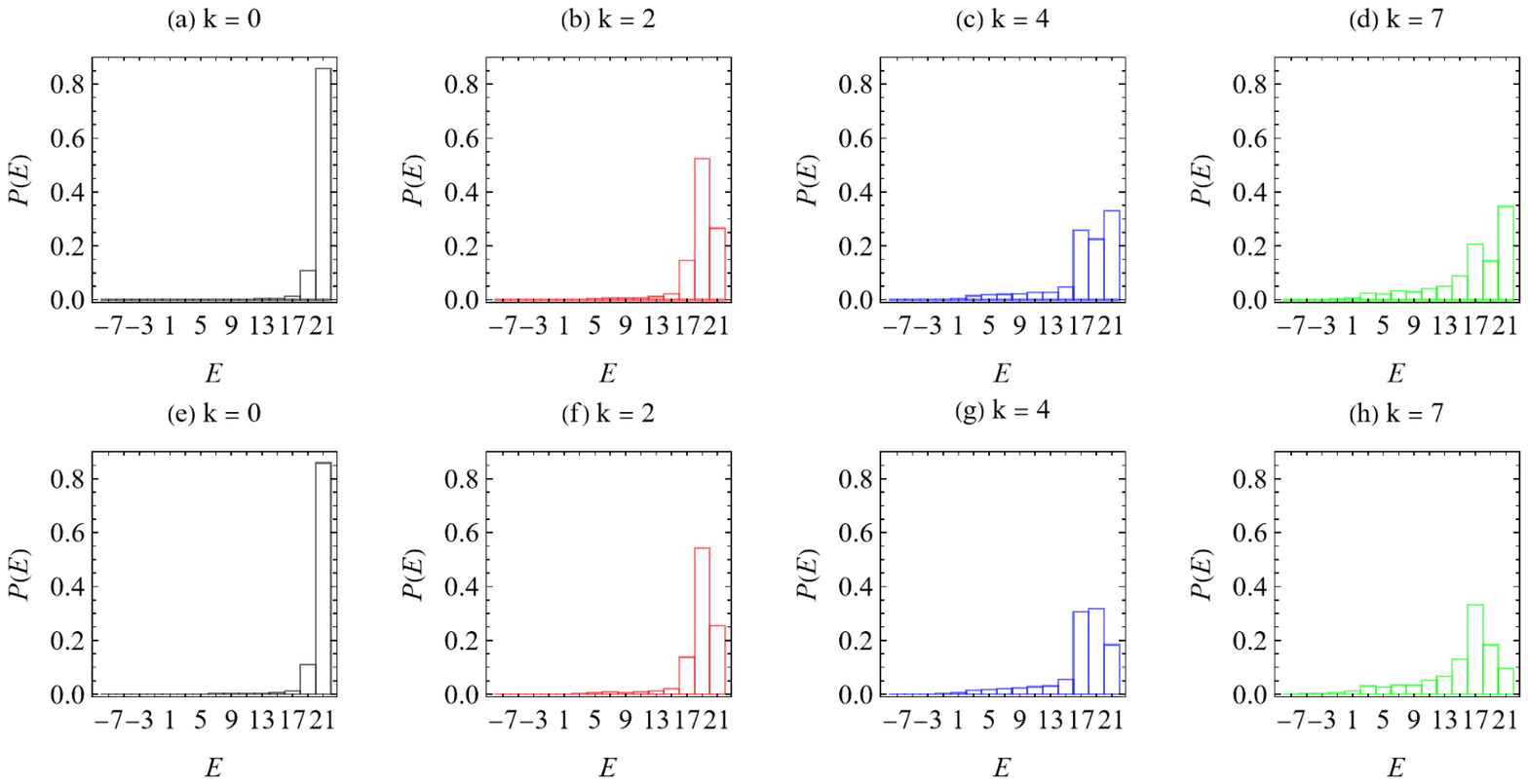} 
\caption{\small Sum of square modulus of coefficients $c_{k} \equiv \langle E_{k} | \psi_0 \rangle$ in each interval of two units of energy for cases (a) (first row) and (b) (second row), $\ket{\psi_0}$ with lack of resolution in the magnetization in the $z$ direction, and Hamiltonian \ref{Hamiltonian}, with $J_{y} = 1.4 \, s^{-1}$, $J_{z} = 0.5 \, s^{-1}$, $h_{z} = 0.01 \, s^{-1}$ and $n=15$.}
\label{espectro}
\end{figure*}

\begin{figure}[!th]
\includegraphics[scale=0.8]{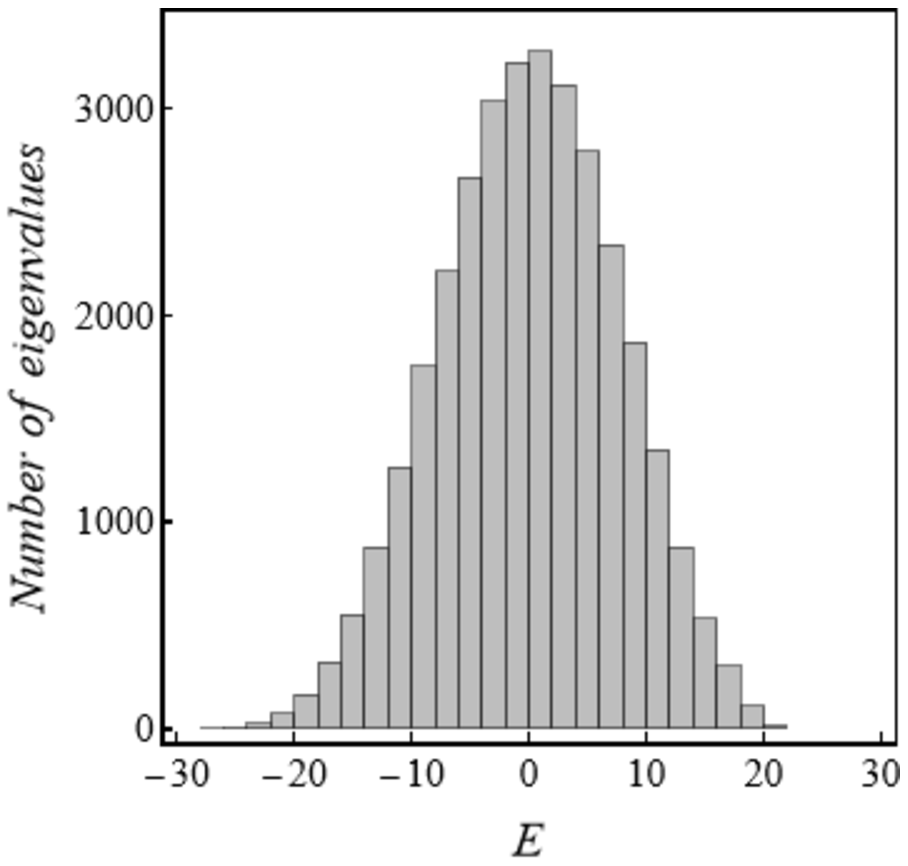} 
\caption{\small Histogram of energy eigenvalues for Hamiltonian \ref{Hamiltonian}, with $J_{y} = 1.4 \, s^{-1}$, $J_{z} = 0.5 \, s^{-1}$, $h_{z} = 0.01 \, s^{-1}$ and $n=15$. Each bar has a width of two energy units. The vertical axis is the number of eigenstates in each interval.}
\label{Heigen}
\end{figure}

In Fig. \ref{espectro} we show the LDOS for cases (a) and (b) as a function of the energy for $n=15$ and four different sets of $k$-flipped spins. For other values of $n$, the behavior is similar. Each bar corresponds to an interval of two energy units, and the frame ticks' numbers are the median values. The height of the bars is calculated by adding the quantities $|c_{k}|^2$ present in each energy interval. We also plot the energy spectrum for Hamiltonian (Eq. \ref{Hamiltonian}) in Fig. \ref{Heigen} showing, as expected, a Gaussian shape centered at zero. This is also useful to visualize the
position of the LDOS in the spectrum of $H$. From Fig. \ref{espectro}, we can see that an initial state with larger coarse-graining has a more broad LDOS, that allows for larger $d_{eff}$ and better equilibration, as expected from the upper bound in Eq. \ref{equilibr}. Comparing Figs. \ref{Heigen} and \ref{espectro} we can see that the initial state without any coarse-graining is concentrated in the border of the spectrum and with a very peaked LDOS and small $d_{eff}$. 

We should mention that we found the same behavior of the effective dimension in Figure \ref{EdxD} for some other values of the parameters of $H$ in the non-integrable regime. It is also worth noting that while coarse-graining is incorporated in the preparation and description of the initial state, it is not applied to the observable. Thus, while the initial state may have a coarse resolution, we can precisely measure the magnetization observable \ref{observable} in the situation under consideration.

%From eq. \ref{deff} we see that if we have a superposition of lots of energy eigenstates as an initial state, we are contributing to equilibration. However, if we have an initial state with definite energy, the denominator in \ref{deff} will only have one term and equilibration won't be the case. That is why macroscopic systems generally equilibrate: the distance between the energy levels is small, making it difficult to prepare initial states localized in the energy spectrum \cite{Reimann2008}. It is also known \cite{PRLLuis} that a system on the edge of the energy spectrum does not equilibrate. It should be stressed that our initial state is written on the (local) magnetization basis. To gain information about the distribution of our initial state in the Hamiltonian spectrum, we need to evaluate $c_{k}$.

%Given the energy spectrum of the Hamiltonian, and comparing it together with Fig. \ref{espectro}, we see that the less the number of $k$-flipped spins, the more located at the edge of the spectrum are the coefficients $c_{k}$. As the number of spins flipped down increases, more coefficients $c_{k}$ occupy intervals further from the edge. This was the expected result, given that energy distributions with most of its coefficients located at the edge disfavor equilibration. Therefore, Fig. \ref{espectro} shows us that our coarseness in resolving the initial state favors equilibration.

\section{Equilibration time and Coarse-graining}
\label{sec:eqcg}

The upper bound in Eq. \ref{equilibr} gives sufficient conditions for a system to equilibrate, but, as infinite time averages are used, there is no information about the time scales for the equilibration to happen.
Furthermore, if the time scales involved are very large, the results are not relevant to explain the rapid equilibration we observe in nature. Most numerical simulations show reasonable equilibration times, but there is still no consensus about general proofs. If one considers finite time averages, for example, the equilibration time obtained increases with the smallest energy gap, and, therefore, exponentially with the system size for local Hamiltonians \cite{Short12}. Some works are trying to obtain an equilibration time: see \cite{Wilming18}, supplementary information section in \cite{Reimann16} and introduction in \cite{ThiagoNJ} for a brief survey of the literature. Here we will use the approach proposed by one of us and collaborators in \cite{ThiagoNJ} to verify the order of magnitude of the equilibration time when coarse-graining, in the $z$ direction, is present and if the resolution could influence the equilibration time. Let’s consider the time evolution of the relative
fluctuation, given by
\begin{equation}
g(t) = \frac{M_z(t)-\overline{M_z}}{\Delta M_{z}} = \sum_{i \ne j} \frac{(c_{j}^{*} M_{ij} c_{i})}{2} e^{-i(E_{i}-E_{j})t} = \sum_{\alpha} \nu_{\alpha} e^{i G_{\alpha} t},
\label{timesignal}
\end{equation} 
with $M_{z}(t)=\< \hat{M_{z}}(t)\>$, $\overline{M_{z}}$ its infinite-time average, $\Delta M_{z}$ the difference between the highest and lowest eigenvalues,  $G_{\alpha} = G_{(i,j)} = E_{j} - E_{i}$ the energy gaps, $\nu_{\alpha} = \nu_{(i,j)} = c_{j}^{*} M_{ji} c_{i}/2$ and $M_{ij}=\<E_i|\hat{M}_z|E_j\>$. 

The main idea to obtain an estimate of the equilibration time is to note that the size of the fluctuation is the modulo of a sum of complex numbers. If one considers each complex number as a vector in the complex plane, it is clear that an initial out-of-equilibrium $|\psi_0\rangle$  has similar complex number phases, such that the sum does not cancel out. But, as time evolves, each number gets a different phase and the terms in the sum start to cancel each
other. This is the usual dephasing mechanism, and an estimate for the dephasing time is the time for the initial phases to spread around $2\pi$. Based on these arguments, an estimative for the equilibration time is given by \cite{ThiagoNJ}
\begin{equation}
T_{eq} \sim \pi / \sqrt{\sum_{\alpha} q_{\alpha} G_{\alpha}^2},
\label{Teq}
\end{equation} 
with $q_{\alpha}:= |\nu_{\alpha}|^2/\sum_{\beta} |\nu_{\beta}|^2 $. The denominator in \ref{Teq} is the dispersion of the distribution of the energy gaps weighted by the $c_k$ and $M_{ij}$ (to take into account the role of the initial state and observable). Note that if the distribution of the $v_\alpha$ is smooth, once the dephasing occurs, the phases spread around $2\pi$, and there are only small fluctuations. Revivals will occur for finite systems, but at scales that grow exponentially with the size of the system.

\begin{figure}[!th]
\includegraphics[scale=0.8]{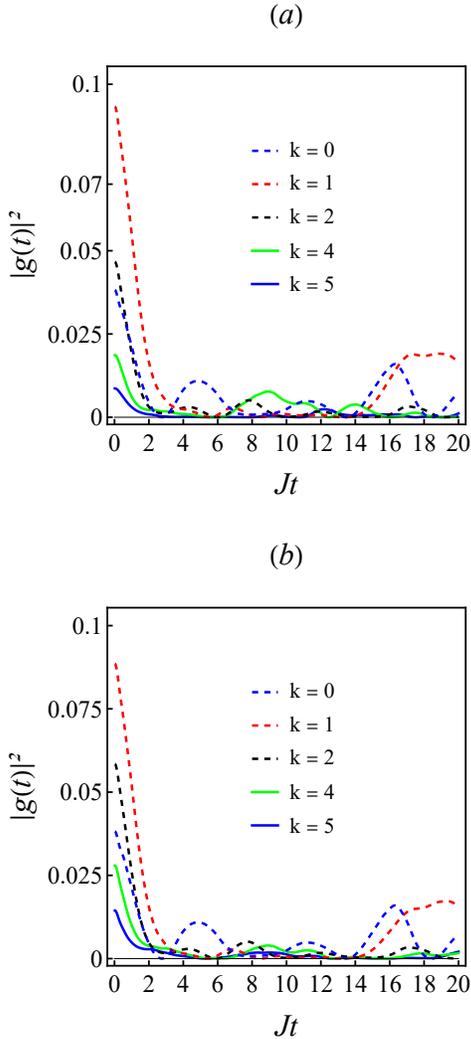} 
\caption{\small Time evolution of the relative fluctuation \ref{timesignal} for case (a) and case (b), initial states with lack of resolution in the magnetization in the $z$ direction, and Hamiltonian \ref{Hamiltonian}, with $n=13$, $J_{y} = 1.4 \, s^{-1}$, $J_{z} = 0.5 \, s^{-1}$ and $h_{z} = 0.01 \, s^{-1}$.}
\label{TimeSignal}
\end{figure}

\begin{figure}[!th]
\includegraphics[scale=0.8]{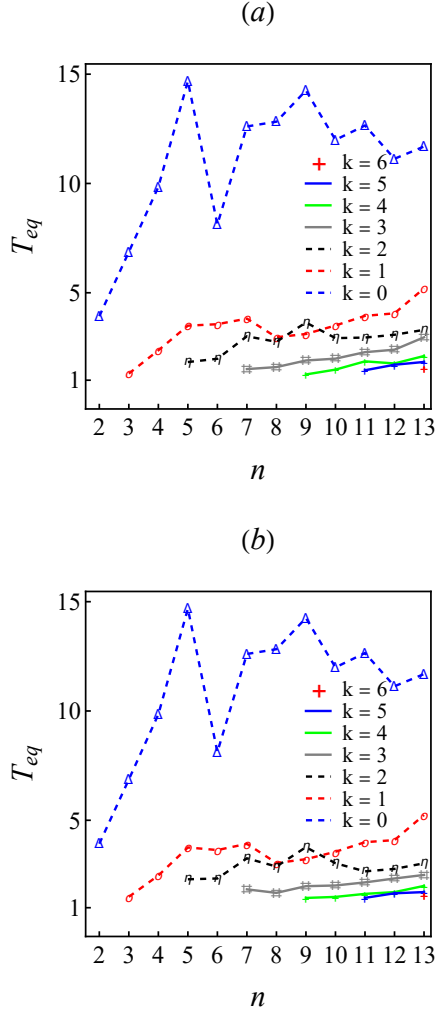} 
\caption{\small Equilibration time versus $n$ for case (a) and case (b), initial states with lack of resolution in the magnetization in the $z$ direction, and for Hamiltonian \ref{Hamiltonian}, with $J_{y} = 1.4 \, s^{-1}$, $J_{z} = 0.5 \, s^{-1}$ and $h_{z} = 0.01 \, s^{-1}$.}
\label{eqtime}
\end{figure}

In Fig. \ref{TimeSignal} we show the square modulus of the time evolution of the relative fluctuation $g(t)$ for both cases (a) and (b) for $n=13$ and different values of parameter $k$. We observe the expected behavior of fluctuations decaying fast to zero and then staying close to it with only small deviations; the revivals can not be seen, but their scale is exponentially large with the system size. Fig. \ref{eqtime} shows the estimates for equilibration times computed via \ref{Teq} for cases (a) and (b) as a function of $n$, the number of spins. The estimate $T_{eq}$ seems to decrease with $k$, but this is only weak evidence since this is an estimate and not a very precise prediction and all the values have almost the same order of magnitude. We can also see that the values obtained are of the same order of magnitude of the time that $g(t)$ gets close to zero. Therefore, we show that the system equilibrates in a finite time and there is an indication that the coarse-graining can also influence the equilibration time. 

\section{Conclusions and Perspectives}

There is a renewed interest in the study of equilibration and thermalization of isolated quantum systems. In this scenario, the system may appear to be at equilibrium because most of the time the fluctuation of some observables is too small to be measured. Rigorous theorems show the conditions for a Hamiltonian, observable, and initial state to equilibrate: the observable should be local, $H$ should not contain many degenerated gaps and the initial state has to be in a superposition of a large number of energy eigenstates. Therefore, to prepare an initial state that does not equilibrate one needs an apparatus with an excellent resolution for energy measures. Thus, one can interpret the lack of resolution in the energy measure/preparation, or a coarse-graining measure/preparation, as the origin of equilibration. On the other hand, many numerical results show that a typical spins chains Hamiltonian with initial states that are products in the local bases equilibrate.

In this work, we studied how a lack of resolution in the magnetization of the initial state affects equilibration. 
We studied two kinds of coarseness: (a) spatial coarse-graining, in which we do not have the spatial resolution to measure the magnetization of particular spins; (b) a magnetization coarse-graining, in which we have a lower bound in the initial state's magnetization spectrum. For a chaotic spin chain, we found that more coarse-grained initial states in the magnetization in $z$ direction have a higher tendency for equilibration; with smaller fluctuation around the equilibrium. We also studied how these magnetic coarse-grained states are decomposed in the energy bases, showing that more coarse-grained states in the $z$ direction have a more delocalized distribution in the energy bases. This agrees with rigorous bounds that connect equilibration with the level of coarse-graining in the energy. We also check for initial states with coarseness in magnetization in other directions than $z$, namely $x$, $y$ and the direction given by the angles $(\theta=30\degree, \phi=90\degree)$. We can see that the lack of resolution in the magnetization affects the tendency for equilibration, but not always increasing it. Finally, we studied how fast equilibration happens, showing that the estimates proposed in \cite{ThiagoNJ} do predict the order of magnitude of the equilibration time. We also found some evidence that the equilibration time may decrease with the level of coarse-graining. 

From perspectives, it would be interesting to study deeply the effects of coarse-graining in other models, in other magnetization directions, and to consider other kinds of coherent superpositions besides the "uniform" one used here: we considered a uniform superposition of all possible states with a well-defined magnetization as the coarse-grained initial states. A natural possible next step is to put a different probability distribution in the weights of the superposition to see if it is possible to identify changes in the results, since there are examples where they are relevant \cite{gemmercoef}. Or even treat the initial states as statistical mixtures instead of coherent sums. Another interesting point is to study the nature of the equilibrium state: if it is thermal, for example, and the relation between coarse-graining and ETH, which are known to be related \cite{gemmerETH}.

%Part of the motivation for this work was to study macroscopic superpositions, trying to understand that we do not see macroscopic superpositions in our daily lives. As already commented, one reason may be the proximity of the energy levels \cite{Reimann2008}, which would cause them to equilibrate very quickly. In \cite{macrosuper} the authors argue that distinguishing macroscopic superpositions from incoherent mixtures requires a quadratically bigger measurement device than the superposition state.

%Being possible or not to construct such a device, our line of thought resides on the fact that we need to describe macroscopic superpositions at a coarse-graining level since it is practically impossible to have access to all its degrees of freedom. Maybe this coarse description is one of the responsible for a very fast equilibration time.

%%%%%%%%%%%%%%%%%%%%%%%%%%%%%%%%%%%%%%%%%%%%%%%%
%%%%%%%%%%%%%%acknowledgments%%%%%%%%%%%%%%%%%%%%%%%%%%
%%%%%%%%%%%%%%%%%%%%%%%%%%%%%%%%%%%%%%%%%%%%%%%%

\begin{acknowledgments}
This work is supported by the Instituto Nacional de Ciência e Tecnologia de Informação Quântica (465469/2014-0), by the Air Force Office of Scientific Research under award number FA9550-19-1-0361, and by the National Council for Scientific and Technological Development, CNPq Brazil (projects: Universal 406499/2021-7, and 409611/2022-0). We also thank the referee for the insightful observations and valuable feedback.
\end{acknowledgments}

\end{document}